\documentclass{nature_fig}

\usepackage{graphicx,dcolumn,bm,epic, eepic,fleqn,float}
\usepackage{amssymb,amsmath,amsfonts,multirow,rotate,color}
\usepackage{soul}
\usepackage{wasysym}
\usepackage{subfigure}

\usepackage{hyperref}
\title{Reducing measles risk in Turkey through social integration of Syrian refugees}

\author{Paolo Bosetti$^{1,2\star}$, Piero Poletti$^{1\star}$, Massimo Stella$^1$, Bruno Lepri$^1$, Stefano Merler$^{1\dagger}$ \& Manlio De Domenico$^{1\dagger}$}

\begin{document}
\maketitle
\begin{affiliations}
 \item Fondazione Bruno Kessler, Via Sommarive 18, 38123 Povo (TN), Italy
 \item Department of Mathematics, University of Trento, Trento, Italy\\
$^\star$ These authors contributed equally to this work\\
$^\dagger$ Joint senior authors of this work
\end{affiliations}
\begin{abstract}
Turkey hosts almost 3.5M refugees and has to face a humanitarian emergency of unprecedented levels.
We use mobile phone data to map the mobility patterns of both Turkish and Syrian refugees, and use these patterns to build data-driven computational models for quantifying the risk of epidemics spreading for measles –- a disease having a satisfactory immunization coverage in Turkey but not in Syria, due to the recent civil war 
–- while accounting for hypothetical policies to integrate the refugees with the Turkish population. Our results provide quantitative evidence that policies to enhance social integration between refugees and the hosting population would reduce the transmission potential of measles by almost 50\%, preventing the onset of widespread large epidemics in the country. Our results suggest that social segregation does not hamper but rather boosts potential outbreaks of measles to a greater extent in Syrian refugees but also in Turkish citizens, although to a lesser extent. This is due to the fact that the high immunization coverage of Turkish citizens can shield Syrian refugees from getting exposed to the infection and this in turn reduces potential sources of infection and spillover of cases among Turkish citizens as well, in a virtuous cycle reminiscent of herd immunity.
\end{abstract}

\section*{Introduction}

Human migration represents a complex phenomenon influencing in several inter-connected ways the economy, the healthcare and the social cohesion of whole countries~\cite{abubakar2018ucl,sjaastad1962costs,cheong2007,zimmerman2011health,castles2013age,rechel2013,blitz2017health}. However, it is only recently that the availability of massive datasets opened to new advancements in modelling and understanding such complexity~\cite{lu2012pnas,deville2014pnas,blondel2015survey,lu2016flowminder}, addressing the urgent need for effective, large-scale intervention policies towards managing the consequences of massive migration flows~\cite{castles2013age}. 

In our manuscript, we focus our attention on Turkey, a country facing a humanitarian emergency of unprecedented levels~\cite{UNHCR_annualReport}. In the last eight years, more than 3.5M Syrians, displaced by the war, have sought refuge in Turkey. This number, through births and new arrivals, is also increasing by approximately 1,000 people per day. The arrival of a huge amount of people with different economic, health, and living conditions, and from a country where the healthcare system has been almost completely disrupted, may raise serious concerns about the risks of Turkish health systems being overburdened. 

For instance, Turkish infectious disease specialists are concerned that Syrian refugees' crisis may impose serious risks to their country for infectious diseases previously eliminated or in the process of being eliminated~\cite{hargreaves2016}. 
According to the latest reports from WHO and UNICEF~\cite{who-coverage}, immunization coverage in Syria dropped from more than 80\% before the war to a worrying 41\% in 2015 for the most basic vaccines, resulting in millions of unvaccinated children. Direct consequences of this alarming situation are a high risk of epidemic outbreaks (e.g., evidence for polio~\cite{whopolioreport} and measles~\cite{whomeaslessyria} has been reported) and a potential increase of mortality due to diseases which could be prevented with vaccines~\cite{simons2012assessment}.
Thus, countries, such as Turkey, Lebanon, and Jordan, hosting a great concentration of Syrians perceive the lack of an appropriate immunization coverage as a potential risk of epidemic outbreaks for the local population~\cite{hargreaves2016}. This perceived risk may ignite a cascade of social dynamics which could: i) reinforce the segregation of refugees; ii) increase unemployment and poverty; iii) result in difficult relationships between healthcare workers and Syrians.

The aim of this study is to quantify the risk of observing widespread measles epidemics in Turkey, showing potential public health benefits coming from social integration between Syrian refugees and Turkish citizens. To such aim we developed a transmission model to investigate the influence of social mixing and integration among Syrian refugees and Turkish citizens on the potential spread of measles epidemics in Turkey. The model takes explicitly into account empirical mobility patterns, as inferred by mobile phone data~\cite{salah2018data}, and the current level of immunity against measles in the two considered populations, as estimated from available epidemiological evidences~\cite{who-outbreakDataUsed,who-coverage}. Since the current amount of integration is difficult to estimate with available data, measles transmission is modelled by considering a tunable parameter that accounts for a variety of scenarios, ranging from full segregation to full integration. Results are obtained by simulating the spread of measles by assuming different scenarios for measles transmission potential and different levels of social integration.

In particular, measles represents an illustrative case of a highly contagious infectious disease which can be prevented with a safe and effective vaccine~\cite{grais2006estimating,lessler2011maintaining,simons2012assessment,Merler2014,li2017demographic,Trentini2017,poletti2018hidden}. Despite substantial progress towards measles elimination at the global level has been documented, re-emergence of large measles epidemics was observed in the last decade both in low-income and in high-income countries~\cite{Trentini2017}. Moreover, measles epidemiology varies widely across different geographical regions, as a consequence of heterogeneous immunity gaps, generated by sub-optimal immunization activities, in different socio-demographic settings~\cite{Merler2014,Trentini2017,li2017demographic}. 

 The crucial role played by both human mobility~\cite{ferguson2006strategies,colizza2006role,balcan2009multiscale,Merler2010,Merler2011,Zhang2017} and mixing patterns~\cite{mossong2008social,Fumanelli2012,Ajelli2014} in shaping the transmission dynamics of infectious diseases has been widely documented in the literature and represents a key component of realistic modeling aimed at informing public health policies. Thus, human mobility models have been used to map flows of individuals between geographical areas at different scales and to improve the reliability of transmission models of infectious diseases~\cite{ferguson2006strategies,colizza2006role,balcan2009multiscale,meloni2011modeling,bajardi2011human,lima2015disease,gomez2018critical}. In the last years, mobile phone data have been successfully used as a valuable proxy for human mobility~\cite{gonzalez2008,song2010,simini2012universal,tizzoni2014proxies,blondel2015survey,barbosa2018human}.  We capitalize on these works to build, from mobile phone data, a multilayer network~\cite{de2013mathematical,kivela2014multilayer,de2016physics} map of human mobility of Turkish citizens and Syrian refugees in Turkey, and we use this knowledge to develop a computational model for the potential epidemic spread of measles.

The contribution of our work is twofold. On the one hand, we identify the \emph{epidemic risks} associated with measles in Turkey. On the other hand, we investigate how these epidemic risks are affected by policies devised to enhance \emph{social integration} between Syrian and Turkish populations.

\section*{Results}

\subsection{Immunity levels in the two populations.}
Two different immunity levels against measles infection are estimated for the Turkish and the Syrian populations. As measles epidemics have not been recently reported in Turkey, we assume the measles immunity level among Turkish citizen reflects the fraction of immunized individuals among birth cohorts between 2006-2016 through 1$^{st}$ and 2$^{nd}$ dose routine vaccination programs~\cite{who-coverage} (see Supplementary Methods). Accordingly, our estimates suggest that only 3.8\% of Turkish people might be currently susceptible to measles infection. A different level of susceptibility in the Turkish population is also considered for sensitivity analysis in the Supplementary Discussion.  

\begin{figure}
\includegraphics[width=\textwidth]{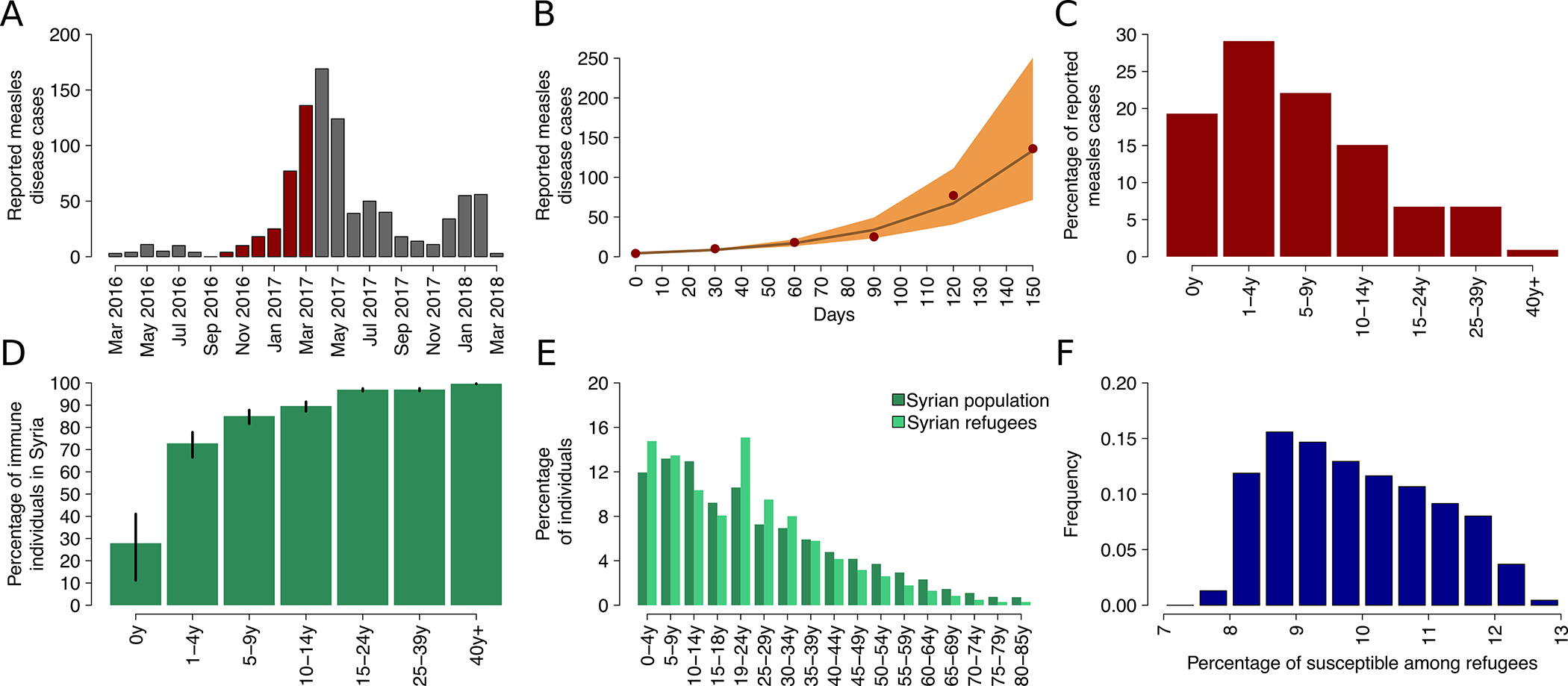}
\caption{\label{fig:estimates}{\bf Measles immunity levels. A} Reported number of measles disease cases over time, during the 2016-2018 measles epidemics in Syria as recently reported by the World Heath Organization~\cite{who-outbreakDataUsed}; red bars correspond to data points used to derive the $R_e$ as a function of the exponential growth rate of the observed epidemic. {\bf B} Obtained fit of the epidemic exponential growth  between September and February 2017 in Syria: red solid line represents the mean estimate, orange shaded area represents 95\%CI. {\bf C} Observed distribution of measles cases across different ages during the 2016-2018 measles epidemics in Syria as recently reported by the World Heath Organization~\cite{who-outbreakDataUsed}. {\bf D} Estimated age specific serological profile in Syria at the beginning of 2017: green bars represents mean values, vertical black lines represent 95\%CI. {\bf E} Observed age distribution of Syrian refugees in Turkey (light green)~\cite{agesyrianrefugees} compared with the population age distribution in Syria (dark green). {\bf F} Estimated percentage of susceptible individuals among Syrian refugees.}
\end{figure}

Estimates of the immunity level among refugees was instead obtained by inferring the age-specific fraction of susceptible individuals in Syria during a recent measles epidemic from the growth rate and age-distribution of cases reported in 2017~\cite{who-outbreakDataUsed}, and accounting for the age distribution of Syrian refugees in Turkey~\cite{agesyrianrefugees} (Fig.~\ref{fig:estimates}). We found that the effective reproductive number ($R_{e}$) of the recent Syrian measles epidemic was 1.32 (95\%CI 1.26--1.38). Consequently, we estimated that the percentage of susceptible individuals in Syria at the beginning of 2017 was 8.92\% (95\%CI 7.29–-10.96). The resulting percentage of susceptible individuals among Syrian refugees in Turkey was estimated to be 9.87\% (95\%CI 8.07-–12.18)(Fig.~\ref{fig:estimates}). 

Obtained results suggest that nowadays, in Turkey, 280,000-430,000 out of 3.5M Syrian refugees and about 3M out of 80M Turkish people are measles susceptible. 

\subsection{Social integration, human mobility and disease transmission.}
\begin{figure}
\includegraphics[width=\textwidth]{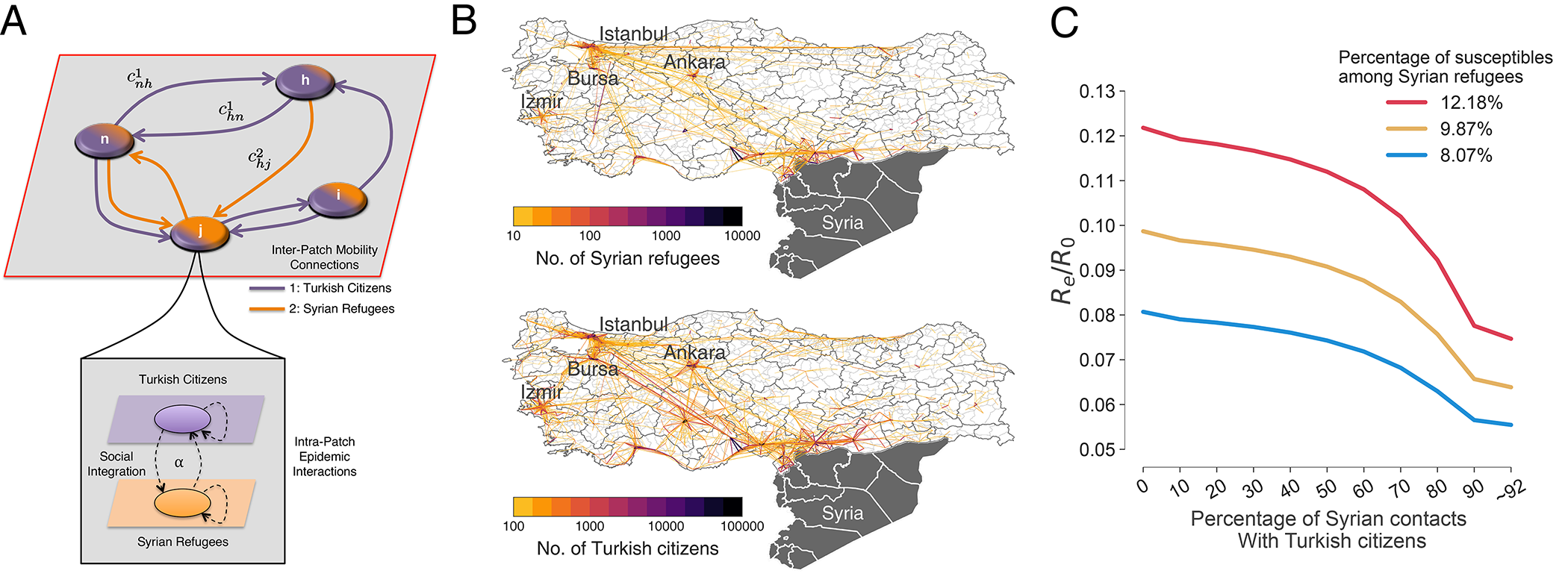}
\caption{\label{fig:Model_Mobility_Re}{\bf Model structure, human mobility and transmission potential. A} Schematic illustration of the model considered in this work. Each prefecture of Turkey is considered as a node of a meta--population network of geographic patches. Two populations, namely Turkish and Syrians, are encoded by different colors and move between patches following the inferred inter-patch mobility pathways. Turkish and Syrian populations encode two different layers of a multilayer system~\cite{de2013mathematical,kivela2014multilayer,de2016physics} where social dynamics and epidemics spreading happen simultaneously. {\bf B} Mobility of Syrian refugees (Top) and Turkish citizens (Bottom) between the prefectures of Turkey as inferred from CDR. Different colors are used to indicate the number of individuals moving from a prefecture to another. {\bf C} Effective reproduction number for measles spreading according to our model, rescaled by $R_0$, as a function of the mixing parameter accounting for social integration between Turkish and Refugees. Colored  lines are associated with the estimated levels of susceptibility among Syrian refugees.}
\end{figure}

The risk of measles re-emergence in Turkey is here analysed by using a compartmental transmission model explicitly taking into account potential infectious contacts occurring between individuals moving across  the country. Different scenarios on how much Syrian refugees interact with Turkish citizens are investigated. A schematic representation of the model is shown in Fig.~\ref{fig:Model_Mobility_Re}A,B along with spatial mobility patterns inferred by the analysis of Call Detailed Records (CDRs) associated with the usage of mobile phones in the country.

The fundamental quantity regulating disease dynamics is the basic reproduction number ($R_0$), which represents the average number of secondary infections in a fully susceptible population generated by a typical index case during the entire period of infectiousness. Larger $R_0$, higher the disease transmissibility. If $R_0>1$ the infection will be able to spread in a population. Otherwise, the infection will die out.  For endemic diseases like measles, $R_0$ provides insights into the proportion $p$ of immune population (either due to vaccination or natural infection) required to prevent large outbreaks; the equation $p=1-1/R_0$ is widely accepted~\cite{anderson1992infectious,grais2006estimating,lessler2011maintaining}. For instance, if $R_0=20$ at least 95\% of the population has to be immune to eliminate the disease. As for measles, typical values of $R_0$ ranges from 12 to 18~\cite{anderson1992infectious,grais2006estimating,lessler2011maintaining,Merler2014,poletti2018hidden}. However, when considering diseases with pre-existing levels of immunity (e.g. childhood diseases like measles), $R_0$ is a theoretical value representing what could happen in terms of disease transmissibility by removing immunity. In these cases, an appropriate measure of diseases transmissibily is provided by the effective reproduction number ($R_e$), which represents the average number of secondary infections in a partly immunized population generated by a typical index case during the entire period of infectiousness.

In Fig.~\ref{fig:Model_Mobility_Re}C we show the ratio $R_e/R_0$ as obtained by varying the fraction of Syrian refugees susceptible to measles from 8.07\% to 12.18\% and by varying the level of social integration from 0\% (full segregation of refugees) to 100\% (full integration of refugees). We found that pre-existing levels of immunity of the two populations reduce $R_e$ to values lower than 10\% of $R_0$. For example, if $R_0$ is lower than 10, the probability of observing an epidemic outbreak would be close to 0 because $R_e$ would result lower than 1 as a consequence of pre-existing immunity levels. However, if $R_0$ is in a more plausible range of values (e.g. 12-18), pre-existing levels of immunity, which are particularly low among Syrian refugees, might not be sufficient to prevent the spread of the disease. Moreover, we found that $R_{e}$ is maximum when the two populations live socially segregated from each other, whereas it quickly decreases by almost 50\% when the two populations are socially well integrated. 

In sum, the immunity level characterizing the Turkish population in 2017 is expected to prevent the spread of future measles epidemic in geographical locations predominantly populated by Turkish citizens. However, if a measles index case would occur in a population with a sufficiently large proportion of Syrian people, transmission events will be sustained by the lack of adequate immunity levels among refugees. Our modelling analysis show that for any scenario considered the risk of observing large epidemics increases with the basic reproduction number and the proportion of susceptible among the refugees (see Fig.~\ref{ARall} and Supplementary Information).

\subsection{Measles epidemic risks in Turkey.}
\begin{figure}
\includegraphics[width=0.8\textwidth]{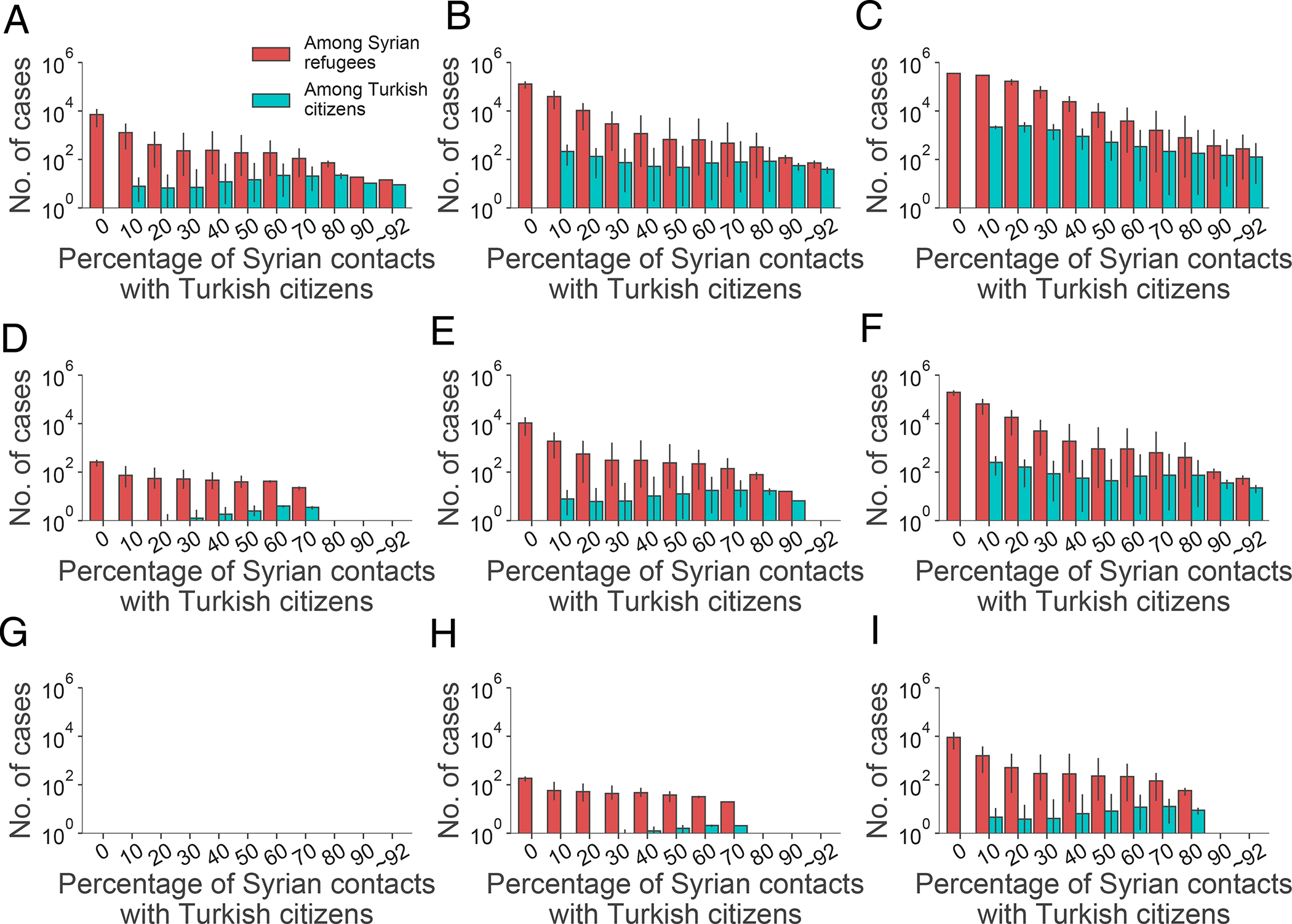}
\caption{\label{ARall} {\bf The potential burden of measles epidemics.} Cumulative infections considering epidemics that exceed 20 cases in the entire population. Bars represent the average number of infections occurring among Syrian refugees (red) and Turkish citizens (blue) for the model projections as a function of the mixing parameter, black lines represent 95\%CI. Three different values of $R_0$ and Susceptibility levels among the Syrian refugees population ($S_0^{Ref}$) were considered. {\bf A} $R_0 = 18$, $S_0^{Ref} = 8.07\%$.  {\bf B} as A but for $S_0^{Ref} = 9.87\%$. {\bf C} as A but for $S_0^{Ref} = 12.15\%$. {\bf D} as A but for $R0 = 15$. 
{\bf E} as B but for $R0 = 15$. {\bf F} as C but for $R0 = 15$.  {\bf G} as A but for $R0 = 12$. 
{\bf H} as B but for $R0 = 12$. {\bf I} as C but for $R0 = 12$.
}
\end{figure}

In case of full segregation of refugees (although practically infeasible and therefore unlikely), potential measles epidemics would result in dramatic health consequences among refugees, causing a huge amount of measles cases widespread in the country (Fig.~\ref{ARall}).  Specifically, when $R_0=15$ is considered and 9.8\% of refugees are assumed to be measles susceptible, the probability of observing an epidemic with more than 20 cases is 100\% (see Supplementary Discussion) and the final size of potential epidemics is expected to exceed 10,000 cases (mean estimate 10,662 95\%CI 3,172--18,414, see Fig.~\ref{ARall}). Our results show that the risk of observing sustained transmission in the country is large for any value of $R_0$ larger than 15 but also for lower values of $R_0$ (e.g. $R_0=12$) if the proportion of refugees susceptible is 9.8\% or more (see Fig.~\ref{ARall} and Supplementary Discussion). 

\begin{figure}
\includegraphics[width=\textwidth]{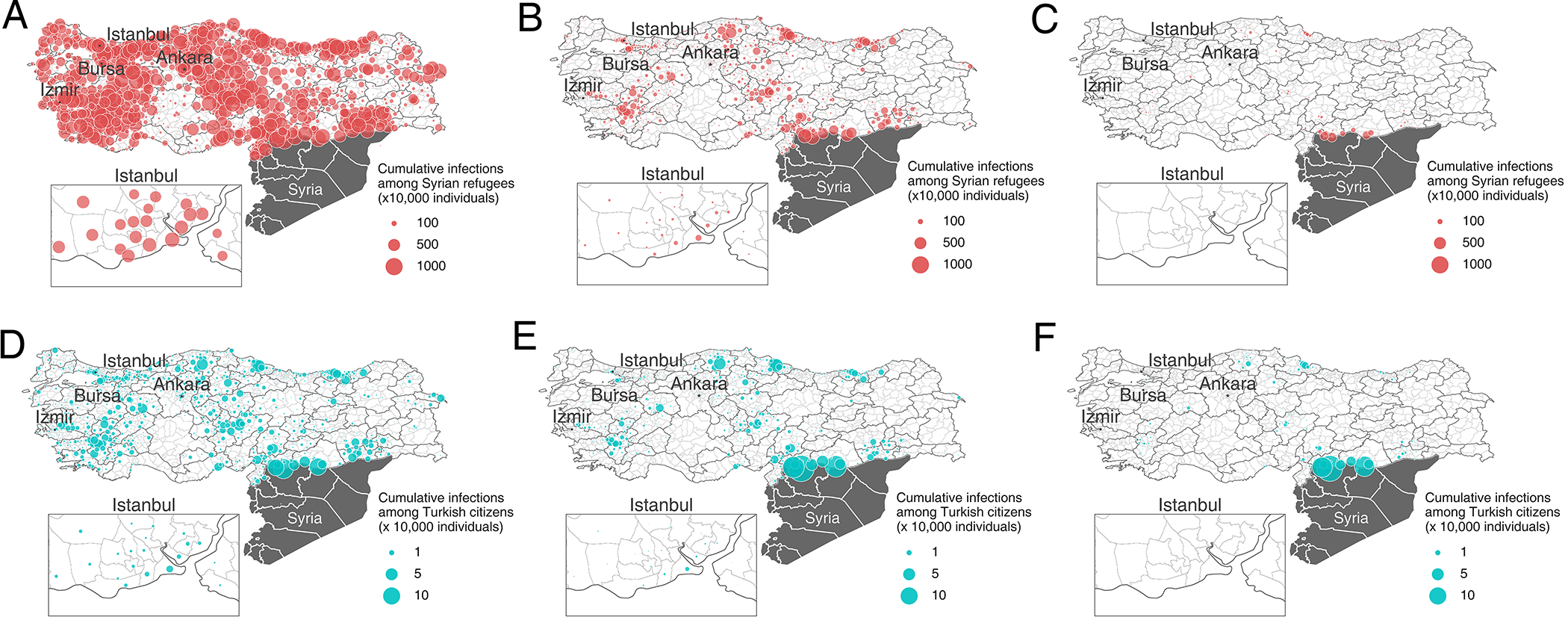}
\caption{\label{Maps_AR}{\bf The potential spatial burden of epidemics. A} shows the estimated cumulative infections in the case of 20\% of Syrian contacts with Turkish citizens considering the worst case scenario in terms of $R_0$ and immunity levels against measles infection among Syrian refugees. Bubbles size are proportional to the average number of measles cases in the Turkish prefectures per 10,000 individuals. Inset displays the Istanbul prefectures. {\bf B} as A but for 40\% of Syrian contacts with Turkish citizens. {\bf C} as A but for 60\% of Syrian contacts with Turkish citizens. {\bf D} as A with respect to the Turkish population. {\bf E} as B  with respect to the Turkish population. {\bf F} as C  with respect to the Turkish population.}
\end{figure}

In the case of full segregation, infections would occur only among refugees. However, when assuming high level of segregation (i.e. only a small fraction, yet equal or greater than 10\%, of refugees’ contacts occur with Turkish citizens), the risk of experiencing large measles outbreak is high (see Supplementary Discussion) and measles epidemics could produce non-negligible spillover of cases among Turkish citizens as well (Fig.~\ref{ARall}). In particular, in a worst case scenario where $R_0=18$, 12.18\% of refugees are susceptible and more than 70\% of Syrian contacts occur with Syrian people, thousands of measles cases are expected all over the country among the Turkish people as well (see Fig.~\ref{ARall}, \ref{Maps_AR}).

More in general, obtained results suggest that the risk of observing sustained measles transmission, the final size of potential epidemics and the populated area at risk of measles infection are significantly smaller in the presence of high levels of integration of refugees (Fig.~\ref{ARall}, \ref{Maps_AR}). 
Specifically, when $R_0=15$ is considered, 9.8\% of refugees are assumed to be measles susceptible and refugees well mix with the Turkish (e.g. more than 70\% of Syrian contacts occur with Turkish people), the probability of observing epidemic outbreak dramatically decreases to values lower than 10\% (see Supplementary Discussion). Moreover, in case of outbreak, the expected overall number of cases is no larger than few hundred (Fig.~\ref{ARall}), as potentially infectious contacts would more probably occur with Turkish immune individuals, who represent about 90\% of individuals currently leaving in Turkey. 

\subsection{Spatial diffusion of potential epidemics.}
\begin{figure}
\includegraphics[width=\textwidth]{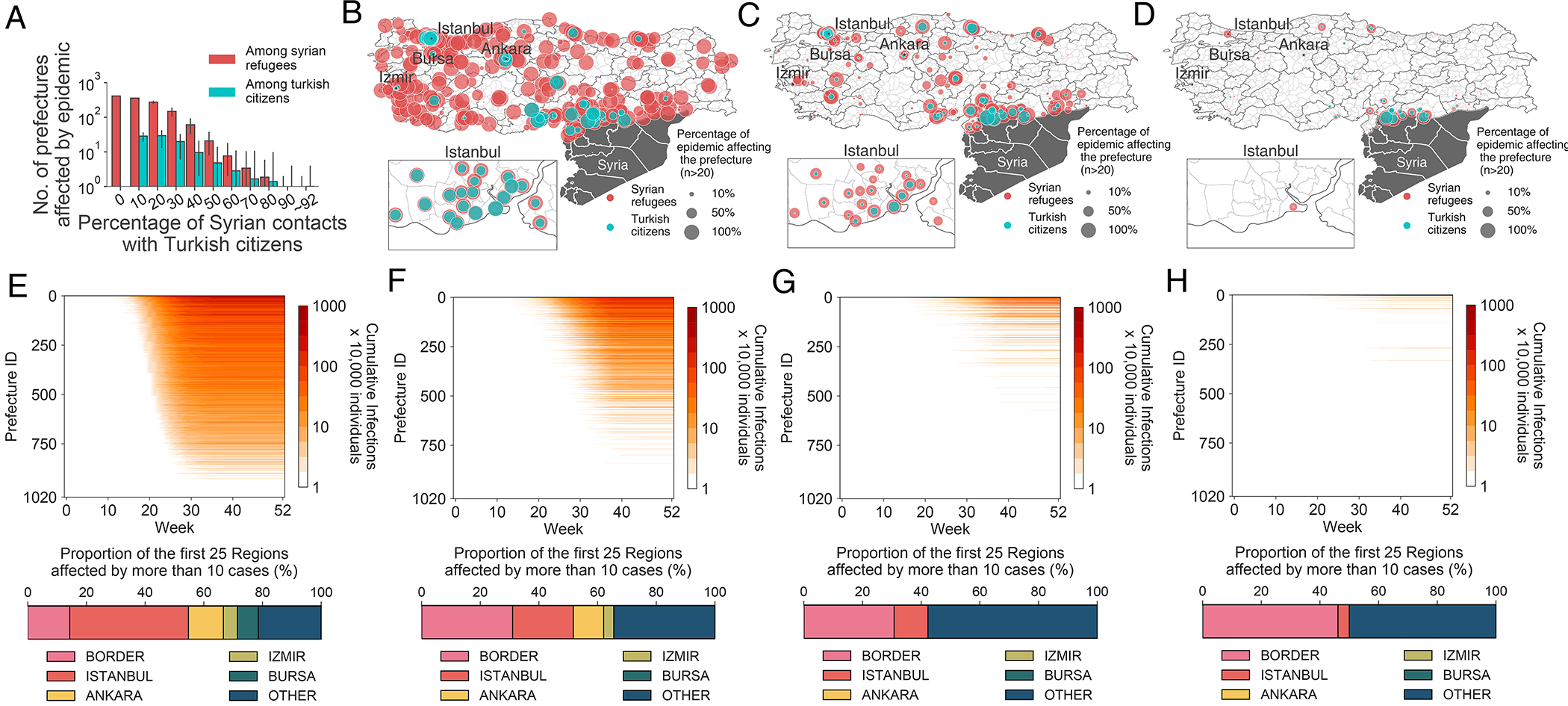}
\caption{\label{Invasion_Panel}{\bf Spatio-temporal spread of potential epidemics. A} shows the estimated number of prefectures affected by the epidemic as a function of the mixing parameter in the worst case scenario. Bars represent the average number of prefectures exceeding 20 cases among Syrian refugees (red) and Turkish citizens (blue); black lines indicate the 95\%CI. {\bf B} Percentage of the simulated epidemic that exceed 20 cases per prefecture in the case of 20\% of Syrian contacts with Turkish citizens. Red and blue bubbles refer to Syrian refugees and Turkish citizens respectively. {\bf C} as B but for 40\% of Syrian contacts with Turkish citizens. {\bf D} as B but for 60\% of Syrian contacts with Turkish citizens. {\bf E} (Top) relative incidence over time considering both the populations per prefecture in the case of total segregation. Prefectures are ranked in decreasing order at week 20. (Bottom) Proportion of region affected in the initial phase of the epidemic considering the first 25 prefectures affected by more than 10 cases. Border refers to Hatay, Kilis, Gaziantep, Sanliurfa, Mardin and Sirnak regions. {\bf F} as E in the case of 20\% of Syrian contacts with Turkish citizens. {\bf G} as E in the case of 40\% of Syrian contacts with Turkish citizens. {\bf H} as E in the case of 60\% of Syrian contacts with Turkish citizens. }
\end{figure}

Remarkably, larger segregation levels also promote the spatial invasion of the epidemic across the whole country (Fig.~\ref{Invasion_Panel}). In the worst case scenario of $R_0=18$, 12.18\% of Syrian refugees susceptible, and more than 90\% of Syrian contacts occur within the Syrian population, the measles epidemic is expected to affect more than 300 out of 1021 prefectures of Turkey (Fig.~\ref{Invasion_Panel}A). On the opposite, if more than 70\% of contacts of refugees would occur with Turkish people, as a consequence of good integration of refugees with Turkish citizens, for the majority of epidemiological scenarios considered, measles epidemics are expected to remain geographically bounded in less than 10 prefectures of the country (Fig.~\ref{Invasion_Panel} and Supplementary Discussion). 

Furthermore, our results suggest that the level of social integration between refugees and the Turkish population can also strongly affect the spatio-temporal spread of potential measles epidemics. Figure~\ref{Invasion_Panel}E-H shows for each prefecture the expected cumulative measles incidence over time for different levels of social integration in the the worst case scenario. Obtained estimates indicate that, in the case of full segregation, 57\% of Turkish prefectures is expected to experience more than 10 measles cases after 30 weeks since the beginning of an epidemic. Such percentage decreases to 11\%, 1\% and 0.4\% when refugees contacts with Turkish citizens increases to 20\%, 40\% and 60\% respectively.
Interestingly, in case of full segregation, prefectures where cases of infections are registered at earlier stages are mainly located in regions associated with the four largest cities of Turkey (64\%, see Figure~\ref{Invasion_Panel}E). In contrast, when 60\% of refugees' contacts occur with Turkish citizens, a remarkable fraction of prefectures that would be affected the early phase of a measles epidemic are close to the Syrian border (Figure~\ref{Invasion_Panel}H). This partially explains why social integration can -- during a potential measles epidemic -- significantly reduce the spillover of cases in the Turkish population.

\section*{Discussion}
The widely accepted critical immunity threshold for measles elimination is 95\% of immune individuals. 
According to our estimates, while Turkish citizens are mostly protected by high vaccine uptake levels, Syrian refugees display a considerably larger fraction of individuals that is susceptibile to the measles, as a consequence of the sub-optimal vaccination during the ongoing civil war. More specifically, while the level of protection of the Turkish population against the disease is nearly optimal (more than 96\% of immune individuals), the protection of Syrian refugees is far from being acceptable (only about 90\% of immune individuals, though highly uncertain).  

The strong difference in the immunity levels among the two populations may have deep repercussions on the society perception towards the movement of Syrians within Turkey. As common in Western countries hosting considerable amounts of migrants~\cite{d2018macroeconomic}, Turkish citizens might perceive the lower immunization coverage of Syrian refugees as a potential threat to national welfare and health. This perception might be even worsened by the staggering numbers of Syrian refugees registered in Turkey, 3.5M in 2018~\cite{salah2018data,d2018macroeconomic}. This well documented negative perception~\cite{ergun2014urban,balkan2015immigration} may trigger segregation mechanisms, aimed at reducing as much as possible interactions and contacts between Syrian refugees and Turkish citizens.

The carried out analysis provides compelling evidence that social segregation does not hamper but rather boosts potential outbreaks of measles to a greater extent in Syrian refugees but also in Turkish citizens, although to a lesser extent.
The main result of the current study is the quantitative evidence that social mixing among Syrian refugees and Turkish citizens can be highly beneficial in drastically reducing the incidence and the strength of infection of measles. This is due to the fact that the high immunization coverage of Turkish citizens can shield Syrian refugees from getting exposed to the infection and this reduces potential sources of infection, in a virtuous cycle reminiscent of herd immunity and well documented in many real-world social systems~\cite{fine1993herd,simons2012assessment}. Our quantitative model combines CDRs data with available epidemiological evidences to estimate the spatial distribution, the immunity profile and the mobility patterns characterizing the two considered populations, allowing the investigation of spatio-temporal patterns of a potential measles epidemic in Turkey.

Provided that a full homogeneous mixing of refugees and citizens could prove to be impracticable or rather difficult to achieve, there are several policies that could reduce social segregation. For instance, designing specific housing policies for redistributing refugees across different neighbourhoods of a given  metropolitan area could avoid the creation of ghettos~\cite{castles2013age}, while also increasing the chances of social interactions between refugees and citizens in schools, shops, third places, etc. 
Although the proposed analysis clearly shows that increasing social mixing between Syrian refugees and Turkish citizens is expected to produce positive public health outcomes, social integration is also expected to provide major societal benefits such as the reduction of violent crimes, economic and educational inequalities~\cite{oecd2016}.

From a geographic perspective, our analysis confirmed that there are metropolitan areas that are pivotal in diffusing the incidence of the disease over time. These areas are mainly prefectures of Istanbul and Ankara and, unsurprisingly, include also many areas adjacent to the national borders of Turkey with Syria. It is in these areas that the efforts for reducing social segregation should be strategically focused. This poses a great challenge for the future, provided that recent reviews of urban regeneration projects highlighted an important process of social segregation of minorities and non-Turkish ethnicity particularly strong in large cities such as Istanbul~\cite{ergun2014urban}. However, our results also suggest that social integration would decrease the relevance of large Turkish cities in promoting the spread of the infection. On the other hand, immunization campaigns targeting areas characterized by a large amount of refugees with respect to the Turkish population, as it is the case of many prefectures close to the Syrian border, might critically reduce the chances of measles transmission and prevent the onset of widespread epidemics.

The performed analysis has several limitations that should be considered in interpreting the results. Estimates of immunity levels in Syrian refugees and in Turkish citizens should be considered cautiously as no recent serological surveys are available for the two populations. Immunity levels are inferred from the analysis of vaccine coverage for Turkish citizens and from the analysis of the 2016-2018 measles outbreak in Syria. This last analysis in particular might be affected by under-reporting of cases~\cite{simons2012assessment,poletti2018hidden} and does not consider potential spatial heterogeneities that could drastically affect estimates of the overall level of protection against the disease. Also, we assume the same levels of immunity in all municipalities, thus neglecting spatial heterogeneities that may be present as a consequence of different vaccine uptake across different regions~\cite{takahashi2017geography}. Moreover, no data on mixing patterns (e.g. by age) are available for either Syrian refugees and Turkish citizens. Consequently, the model neglects potential differences in measles transmissibility by age of individuals and, similarly, potential differences in measles transmissibility for Syrian refugees and Turkish citizens (e.g. induced by different numbers of overall contacts). Finally, CDRs data used in the proposed analysis are associated with only a fraction of the population. Although these data may not perfectly reflect real movements occurring across all the prefectures in the country, they provide valuable evidence to infer a fair approximation of human mobility in the country driving the spatio-temporal spread of an epidemic.

All this considered, the  analysis carried out represents a first attempt to quantify the risk of measles outbreak in Turkey and provides striking evidence that, besides policies aimed at increasing vaccination coverage among Syrian refugees, social integration of refugees within the Turkish population might be an effective countermeasure.

\begin{methods}




The epidemic model used in the our study is an standard one based on three compartments, namely Susceptible, Infected and Recovered (SIR). We refer to the Supplementary Information for details about the model. Here, we provide information about the force of infection used in our model. 

To model the mobility of Turkish and Syrian refugees, we assume two populations of individuals, namely population 1 of size $N^{(1)}$ and population 2 of size $N^{(2)}$, living in a territory consisting of $L$ distinct geographically patches (i.e., Turkish prefectures) accounting for $N_{k}^{(1)}$ and $N_{k}^{(2)}$ individuals, $k=1,...,L$ with $\sum\limits_{k=1}^{L} N_{k}^{(1)} =N^{(1)}$ and $\sum\limits_{k=1}^{L} N_{k}^{(2)} =N^{(2)}$. 

The absolute number of individuals moving between patches is inferred from available Call Detail Records as in Refs.~\cite{lima2015disease,matamalas2016assessing} and rescaled to adequately represent the volumes corresponding to 80M Turkish individuals and 3.5M Syrian refugees. 

Let us indicate by $c_{ki}^{(p)}$ ($p=1,2$) the elements of a matrix $\mathbf{C}^{(p)}$ encoding the number of people belonging to population $p$ travelling from patch $k$ to patch $i$, and with $\alpha$ the fraction of Syrian contacts with Turkish citizens. The force of infection for each population in the $i$--th patch depends on the contribution of all patches in the country: 

\begin{eqnarray}
\lambda_{i}^{(1)}(\alpha,\mathbf{C}^{(1)},\mathbf{C}^{(2)})&=&\beta_{1}\sum_{k=1}^{L}\left[\underbrace{c_{ki}^{(1)}\frac{I_{k}^{(1)}}{N_{k}^{(1)}}}_{\text{Endogenous}}+\underbrace{\alpha c_{ki}^{(2)}\frac{I_{k}^{(2)}}{N_{k}^{(2)}}}_{\text{Exogenous}}\right]\\
\lambda_{i}^{(2)}(\alpha,\mathbf{C}^{(1)},\mathbf{C}^{(2)})&=&\beta_{2}\sum_{k=1}^{L}\left[\underbrace{\alpha c_{ki}^{(1)}\frac{I_{k}^{(1)}}{N_{k}^{(1)}}}_{\text{Exogenous}}+\underbrace{c_{ki}^{(2)}\frac{I_{k}^{(2)}}{N_{k}^{(2)}}}_{\text{Endogenous}}\right],
\end{eqnarray}
where $\beta_{p}=\beta/P_{i}^{(p)}(\alpha,c)$ is the transmission rate for population $p$ and $P_{i}^{(p)}(\alpha,c)$ is an appropriate normalization factor (such that all individuals have the same number of contacts, regardless of geography and citizenship, see SI for further details).  
Each contribution consists of an \emph{endogenous} term, accounting for the infectivity due to individuals from the same population, and an \emph{exogeneous} term, accounting for the infectivity due to the other population. The interplay between mobility and social integration is thus encoded in the force of infection for the two populations.

\end{methods}



\bibliographystyle{naturemag}
\bibliography{sample}


\begin{addendum}
 \item Authors thank Turk Telekom and the other collaborators (e.g. Bogazici University, Tubitak, Data-Pop Alliance, UNICEF, UNHCR, IOM, etc.) of the Data for Refugees Turkey (D4R) challenge as well as the Organization Committee and the Project Evaluation Committee. 
 \item[Competing Interests] The authors declare that they have no
competing financial interests.
 \item[Correspondence] Correspondence and requests for materials
should be addressed to M.D.D.~(mdedomenico@fbk.eu) and S.M~(merler@fbk.eu).
\end{addendum}

\end{document}